\documentclass[referee]{raa}           
\usepackage{graphicx,times}
\usepackage{natbib}
\usepackage{amssymb,amsmath}
\bibpunct{(}{)}{;}{a}{}{,}

\usepackage[a4paper=true,pagebackref=true]{hyperref}
\hypersetup{pdftitle = The title of my PDF, pdfauthor = My name, pdfsubject= The subject, pdfkeywords = keyword1 keyword2 keyword3} 
\hypersetup{colorlinks = true, linkcolor = green, anchorcolor = red, citecolor = blue, filecolor = red, pagecolor = red, urlcolor = red}

\begin{document}

   \title{On the intrinsic shape of gamma-ray spectrum for Fermi blazars
$^*$
\footnotetext{\small $*$ Supported by the National Natural Science Foundation of China  (NSFC: 11763005, etc).}
}

 \volnopage{ {\bf 2012} Vol.\ {\bf X} No. {\bf XX}, 000--000}
   \setcounter{page}{1}

\author{Shi-Ju Kang\inst{1,2,*}, Qingwen Wu\inst{3},  Yong-Gang Zheng\inst{4}, Yue Yin\inst{1}, Jia-Li Song\inst{1}, Hang Zou\inst{1}, Jian-Chao Feng\inst{5,2}, Ai-Jun Dong\inst{5,2}, Zhong-Zu Wu\inst{6}, Zhi-Bin Zhang\inst{6}, Lin-Hui Wu\inst{3}}   

   \institute{ School of Electrical Engineering, Liupanshui Normal University, Liupanshui, Guizhou, 553004, China; {\it kangshiju@hust.edu.cn}\\
               \and
               Guizhou Provincial Key Laboratory of Radio Astronomy and Data Processing;{}\\
               \and
              School of Physics, Huazhong University of Science and Technology, Wuhan, Hubei, 430074, China; {\it qwwu@hust.edu.cn}\\
               \and
              Department of Physics, Yunnan Normal University, Kunming, Yunnan, 650092, China; {\it ynzgy@ynu.edu.cn}\\
               \and               
             School of Physics and Electronic Science, Guizhou Normal University, Guiyang,  550001, China; {fengjc@gznu.edu.cn}\\
             \and
             Department of Physics, College of Science, Guizhou University, Guiyang 550025, China. {}\\
\vs \no
   {\small Received October 10, 2017; accepted February 10, 2018}
}

\abstract{The curvature of the $\gamma$-ray spectrum in blazars may reflect the intrinsic distribution of the emitting electron distribution, which will further give some information on the possible acceleration and cooling processes in the emitting region. The $\gamma$-ray spectra of Fermi blazars are normally fitted either by a single power-law (PL) or a log-normal (call Logarithmic Parabola, LP) form. The possible reason for this differnece is not unclear. We statistically explore this issue based on the different observational properties of 1419 Fermi blazars in the 3LAC Clean sample. We find that the $\gamma$-ray flux (100 MeV-100 GeV) and variability index follow bimodal distributions for PL and LP blazars, where $\gamma$-ray flux and variability index show {a positive correlation}. However, the distributions of the $\gamma$-ray luminosity and redshift follow a unimodal distribution. Our results suggest that the bimodal distribution of $\gamma$-ray flux for LP and PL blazars may be not intrinsic and all blazars may have an intrinsic curved $\gamma$-ray spectrum and the PL spectrum is just caused by the fitting effect due to the less photons. 
}

   \authorrunning{S.-J. Kang et al. }            
   
   \titlerunning{$\gamma$-ray spectral shapes for Fermi blazars}  
   \maketitle

%
\newpage
\section{Introduction}           
\label{sect:intro}

Blazars, including flat-spectrum radio quasars (FSRQs) and BL Lacertae objects (BL Lacs), are most powerful active galactic nucleis (AGNs) with a relativistic jet orientated at a small viewing angle to the line of sight \citep{1995PASP..107..803U}, which show rapid variability and high luminosity, high and variable polarization, superluminal motions, core-dominated non-thermal continuum and strong $\gamma$-ray emissions, etc \citep[]{2002PASJ...54..159Z,{2010ApJ...716...30A},2013A&A...554A..51G,2016ApJS..226...20F,1995ARA&A..33..163W,2005A&A...442...97A,2005AJ....130.1418J}. The multi-wavelength spectral energy distribution (SED) from the radio to the $\gamma$-ray bands of blazars dominantly come from the non-thermal emission, where the SED normally exhibits a two-hump structure in the $log{\nu}-log{\nu}{\rm F}_{\nu}$ space. Location of the peak for the lower energy bump in the SED, $\nu^{\rm S}_{\rm p}$, blazars are used to classify the sources as low (LSP, e.g., $\nu^{\rm S}_{\rm p}<10^{14}$Hz), intermediate (ISP, e.g., $10^{14}\rm Hz<\nu^{\rm S}_{\rm p}<10^{15}$Hz) and high-synchrotron-peaked (HSP, e.g., $\nu^{\rm S}_{\rm p}>10^{15}$Hz) blazars \citep[e.g.,][]{{2010ApJ...716...30A}}. 
BL Lacs show a wide distribution (LSP, ISP and HSP), however, almost all FSRQs are LSP blazars.

It is generally acknowledged that the lower-energy hump is normally attributed to the synchrotron emission produced by the non-thermal relativistic electrons in the jet \citep{1998AdSpR..21...89U}, while the origin of the second hump is still an open issue. In the leptonic model scenarios, the high-energy $\gamma$-rays mainly come from inverse Compton (IC) scattering of the relativistic electrons either on the synchrotron photons inside the jet (synchrotron self-Compton, SSC, process;  e.g., \citealt{1992ApJ...397L...5M,1996ApJ...461..657B,1997A&A...320...19M,2003ApJ...597..851K}) and/or on some other photon populations from outside the jet (external-Compton, EC, process, e.g., \citealt{1992A&A...256L..27D,1993ApJ...416..458D,1994ApJ...421..153S,1996MNRAS.280...67G,1998ApJ...501L..51B}). And the hadron model considers that the high-energy $\gamma$-ray emission  originates from the synchrotron radiation process of extremely relativistic protons (\citealt{2000NewA....5..377A,2001APh....15..121M,2003APh....18..593M,2014MNRAS.442.3026P}), or the cascade process of the proton-proton or proton-photon interactions (e.g., \citealt{1992A&A...253L..21M,1993A&A...269...67M,2000A&A...354..395P,2001PhRvL..87v1102A,2013ApJ...764..113Z, 2014ApJ...783..108C, 2015MNRAS.447.2810Y,2016A&A...585A...8Z}).

The third Fermi Large Area Telescope (LAT) source catalog (3FGL) is now available \citep{2015ApJS..218...23A}. The 3FGL catalog includes 3033 $\gamma$-ray sources: 2192 high-latitude and 841 low-latitude $\gamma$-ray sources, where most sources (1717) belong to blazars \citep{2015ApJ...810...14A}. {Based on the 3FGL \citep{2015ApJS..218...23A}, the third catalog of AGNs detected by the Fermi-LAT (3LAC) is presented \citep{2015ApJ...810...14A}}. The high-confidence clean sample of the 3LAC (3LAC Clean Sample), using the first four years of the Fermi-LAT data, lists 1444 $\gamma$-ray AGNs \citep{2015ApJ...810...14A}, which includes 414 FSRQs ($\sim$~30\%), 604 BL Lac objects ($\sim$~40\%), 402 blazars candidates of uncertain type (BCUs, $\sim$~30\%) and 24 non-blazar AGNs ($<$~2\%). 
It is a good chance to study the nature for $\gamma$-ray emissions of blazars using the such a large sample of Fermi-LAT blazars. The physical properties of blazars have been extensive sample researched based on the Fermi source catalogs (e.g., \citealt {2016ApJS..226...20F,{2016RAA....16..103L},2016RAA....16..173F,{2016ApJS..224...26M},{2016Galax...4...36G},{2014MNRAS.441.3375X}};
{\citealt{2017RAA....17...66L, 2016RAA....16...13C,2012ApJ...753...45S,2015MNRAS.454..115S}}). 
The $\gamma$-ray spectrum of the Fermi-LAT blazars in 3FGL catalog mainly exhibits two different spectral shapes: Logarithmic Parabola (LP) and Power Law (PL) shapes $\gamma$-ray spectrum. The physical origin of the different $\gamma$-ray spectrum is still unclear, which may give some hints on the formation mechanisms of the high-energy electron spectrum in the jet {(e.g., \citealt{1986ApJ...308...78L,2007A&A...466..521T,2009A&A...501..879T,2004A&A...413..489M}; \citealt{2013ApJ...765..122Y}; \citealt{2014ApJ...788..179C}; \citealt{ 2006A&A...448..861M,2009A&A...504..821P,2016A&ARv..24....2M,2016ApJS..226...20F})}. In this work, we aim to find (only focus on) the observational differences for the blazars with LP and PL $\gamma$-ray spectrum through the statistical analyses on the 3LAC Clean Sample. We give some description on the sample selection in Section 2, and the results are shown in Section 3. The  discussion and conclusion are presented in Section 4.

\section{Sample} \label{sec:sample}

The 3LAC Clean Sample lists 1444 $\gamma$-ray AGNs \citep{2015ApJ...810...14A} with 414 FSRQs, 604 BL Lac objects, 402 BCU blazars and 24 non-blazar AGNs. We select 1419 Fermi blazars from 3LAC Clean Sample (3C 454.3 fitted with an exponentially cutoff power law is neglected). In this sample, the spectra of 130 sources are fitted with LP and that of 1289 sources are fitted with PL. In table \ref{tab:data}, we present the 3FGL Fermi names, redshift, optical classifications (BL Lac and FSRQ, or BCUs), $\gamma$-ray spectral shape (LP and PL) for selected blazars from the 3LAC Website version\footnote{http://www.asdc.asi.it/fermi3lac/}. In addition, we also collect their variability index\footnote{Sum of 2 $\times$ log(Likelihood) difference between the flux fitted in each time interval and the average flux over the full catalog interval, see Table 16 in \cite{2015ApJS..218...23A}.}, $\gamma$-ray energy flux (${\rm S}_{\gamma}$, from 100 MeV to 100 GeV obtained by spectrum fitting) from the 3FGL. The $\gamma$-ray luminosity (${\rm L}_{\gamma}$) is calculated with ${\nu}{\rm L}_{\gamma}=4\pi{D^2_{L}}{\nu}{\rm S}_{\gamma,{\rm corr}}$, where $D_{L}$ is the luminosity distance\footnote{Here, we adopt the cosmological parameters $H_{0} = 70 \text{km} \cdot \text{s}^{- 1} \cdot \text{Mpc}^{-1}$, $\Omega_{M}=0.27$, $\Omega_{r}=0$ and $\Omega_{\Lambda}=0.73$.}. The $\gamma$-ray energy fluxes, ${\rm S}_{\gamma}$, of PL $\gamma$-ray spectrum in the source rest frame are K-corrected with the formula ${\rm S}_{\gamma,{\rm corr}}={\rm S}_{\gamma,{\rm obs}}(1+z)^{\alpha-1}$ (e.g., \citealt{2016RAA....16..103L,2016ApJS..224...26M}), where $z$ is the redshift, $\alpha$ is the spectral index ($\alpha=\Gamma_{\rm ph}-1$, {$\alpha$ comes from the $\Gamma_{\rm ph}$ values}, $\Gamma_{\rm ph}$ is power-law photon index, {where $dN/dE=K(E/E_0)^{-\Gamma}$}) \citep{2015ApJS..218...23A}. For LP type of $\gamma$-ray spectrum blazars, the $\gamma$-ray flux are calculated using a modified K-correction according to its spectral shape ${\rm S}_{\gamma,{\rm corr}}={\rm S}_{\gamma,{\rm obs}}(1+z)^{1-\alpha'-\beta{{\rm log}(1+z)}}$ ({see \citealt{2016ApJS..224...26M} for the details and references therein}), $\beta$ is curvature parameter,  {$\alpha'$ is obtained from fitting the $\gamma$-ray spectrum using a log parabola instead of the usual power law, where $dN/dE=K(E/E_0)^{-\alpha'-\beta logE/E_0}$} \citep{2015ApJS..218...23A}. In this sample, there are 759 sources with the measured redshift (109 LP sources and 650 PL sources). The $\gamma$-ray luminosity are also listed in Table \ref{tab:data}.

\begin{table*}
	\centering
	\caption{Data of sample}
	\label{tab:data}
	\begin{tabular}{ccccccc}  
		\hline
		  {3FGL name} &   {Optical class}    &   {Spectral type} &   {z}  &   {$\rm S_{\gamma}$}  &   {VI}   &   {log ${\rm L}_{\gamma}$}\\
		\hline
 3FGL J0001.2$-$0748  &	 BL Lac 	        &	 PL     	&	 ... 	         	&	7.82E-12	&	49.74 	&	... 	       \\
 3FGL J0001.4$+$2120 &	 FSRQ      	&	 LP           &	 1.106	 	&	8.07E-12	&	130.34 	&	46.27 	\\
 3FGL J0002.2$-$4152  &	 BCU                 & 	 PL		&       ...                     &      3.02E-12   &	56.35 	&      ...             \\
 3FGL J0003.2$-$5246  &	 BCU                 &	 PL            &	...                      &	3.01E-12   &	45.28 	&      ...             \\
 3FGL J0004.7$-$4740 &	 FSRQ         	&	 PL     	&	 0.880	 	&	9.19E-12	&	112.93 	&	46.67 	\\
 3FGL J0006.4$+$3825 &	 FSRQ        	&	 PL     	&	 0.229	 	&	1.04E-11	&	80.46 	&	45.27 	\\
 3FGL J0008.0$+$4713 &	 BL Lac 	  	&	 PL     	&	 2.100 	 	&	2.19E-11	&	36.71 	&	47.89 	\\
 3FGL J0008.6$-$2340 &	 BL Lac 	  	&	 PL     	&	 0.147	 	&	3.47E-12	&	47.35 	&	44.28 	\\
 \vdots            &    \vdots  &  \vdots   &    \vdots     &  \vdots   &   \vdots      &   \vdots         \\
		\hline
	\end{tabular}\\
 \footnotesize{Note---Columns 1 is the 3FGL names; Columns 2  lists the optical class; Columns 3 gives the gamma-ray spectral type; Column 4 is redshift;  Column 5 gives the energy flux ($\rm S_{\gamma}$~${\rm erg~cm^{-2}~s^{-1}}$); The variability index (VI) is listed in column 6; Columns 7 reports the gamma-ray luminosity (${\rm L}_{\gamma}~{\rm erg~s^{-1}}$) with a measured redshift.}\\
(This table is available in its entirety in xls format.)
\end{table*}

\section{Results} \label{sec:result}
  
\begin{figure*}
  \centering
   \includegraphics[height=6cm,width=7.2cm]  {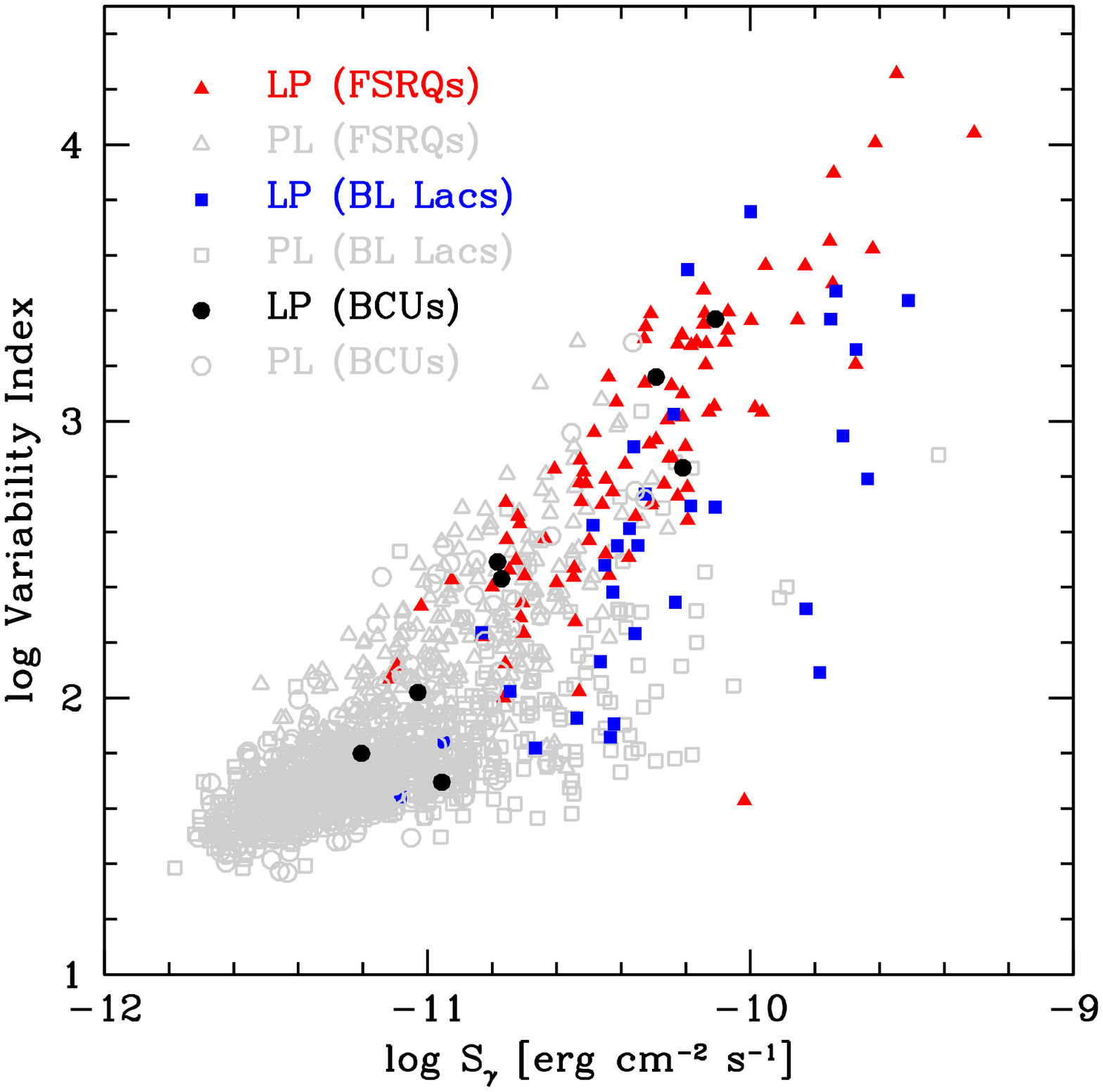}
   \includegraphics[height=6cm,width=7.2cm]  {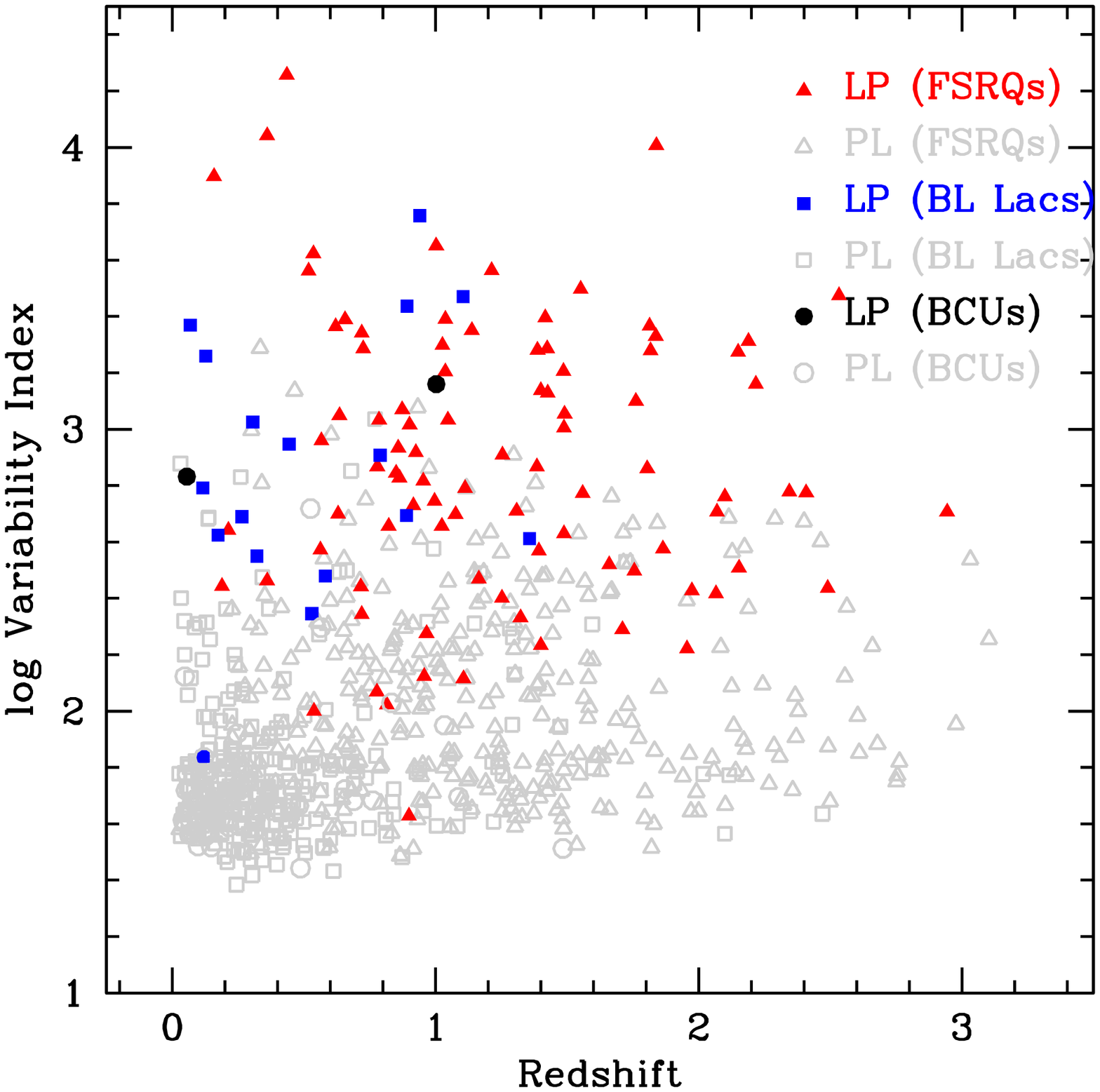}
   \includegraphics[height=6cm,width=7.2cm]  {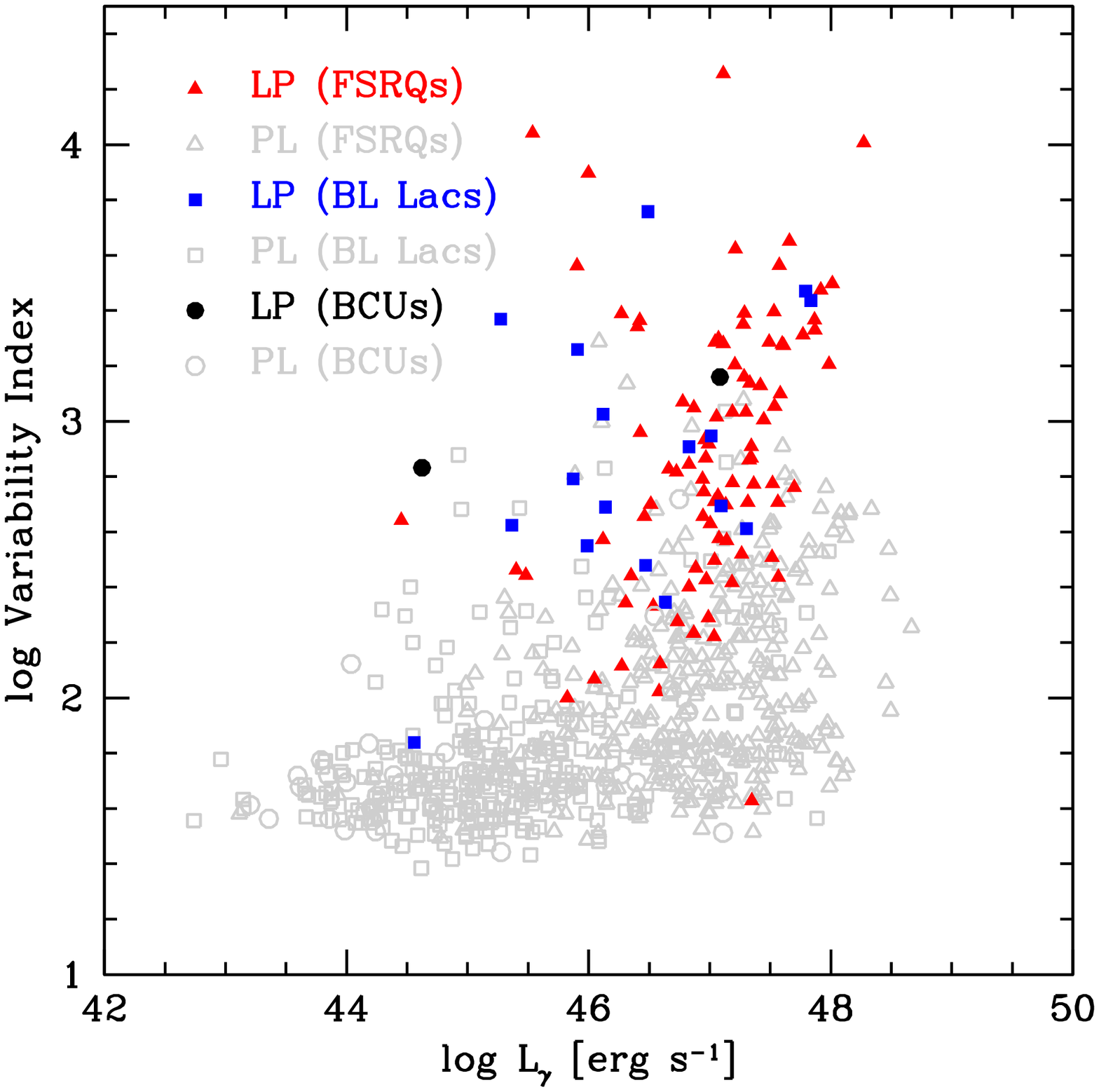}
   \includegraphics[height=6cm,width=7.2cm]  {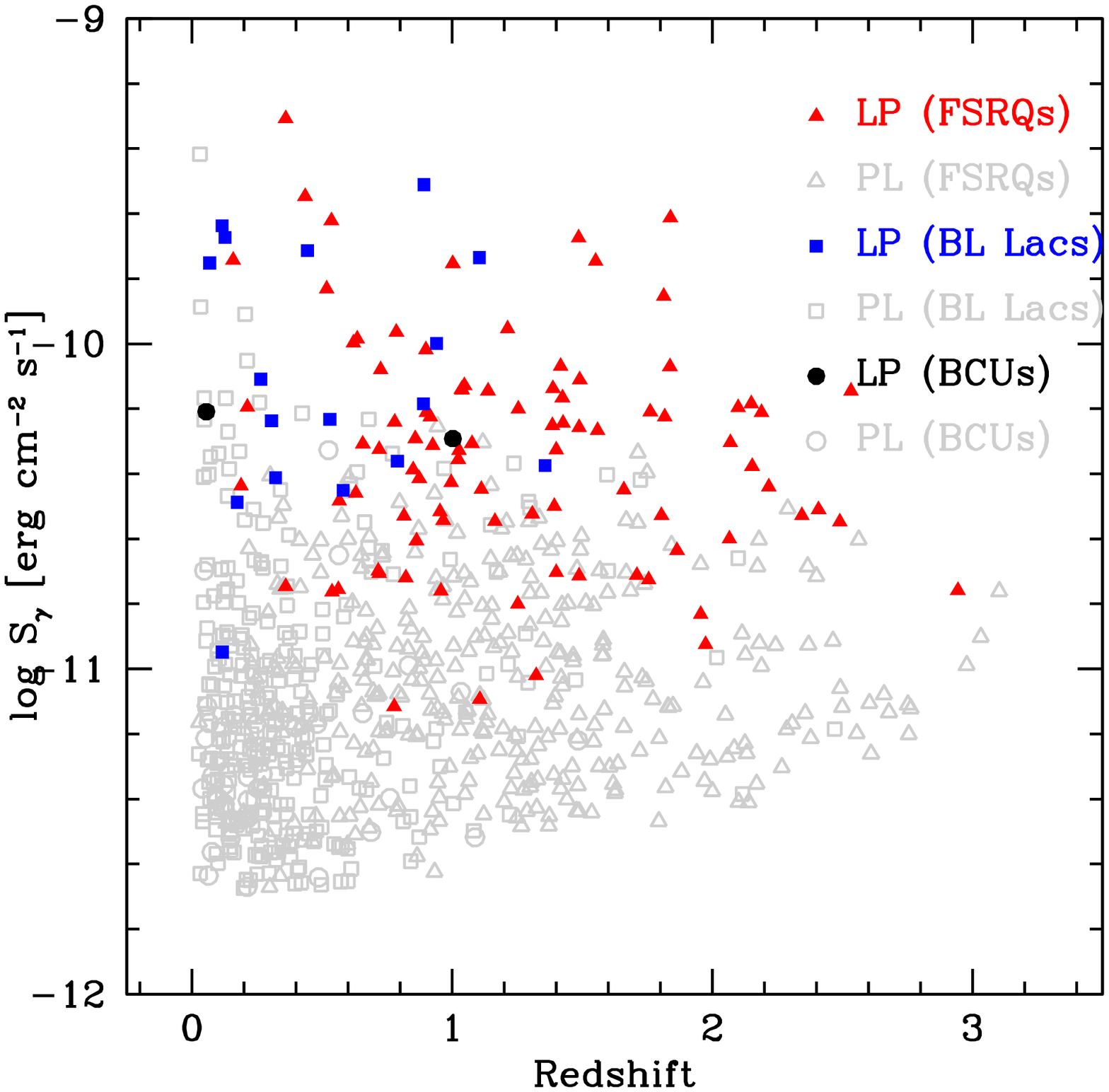}   
   \includegraphics[height=6cm,width=7.2cm]  {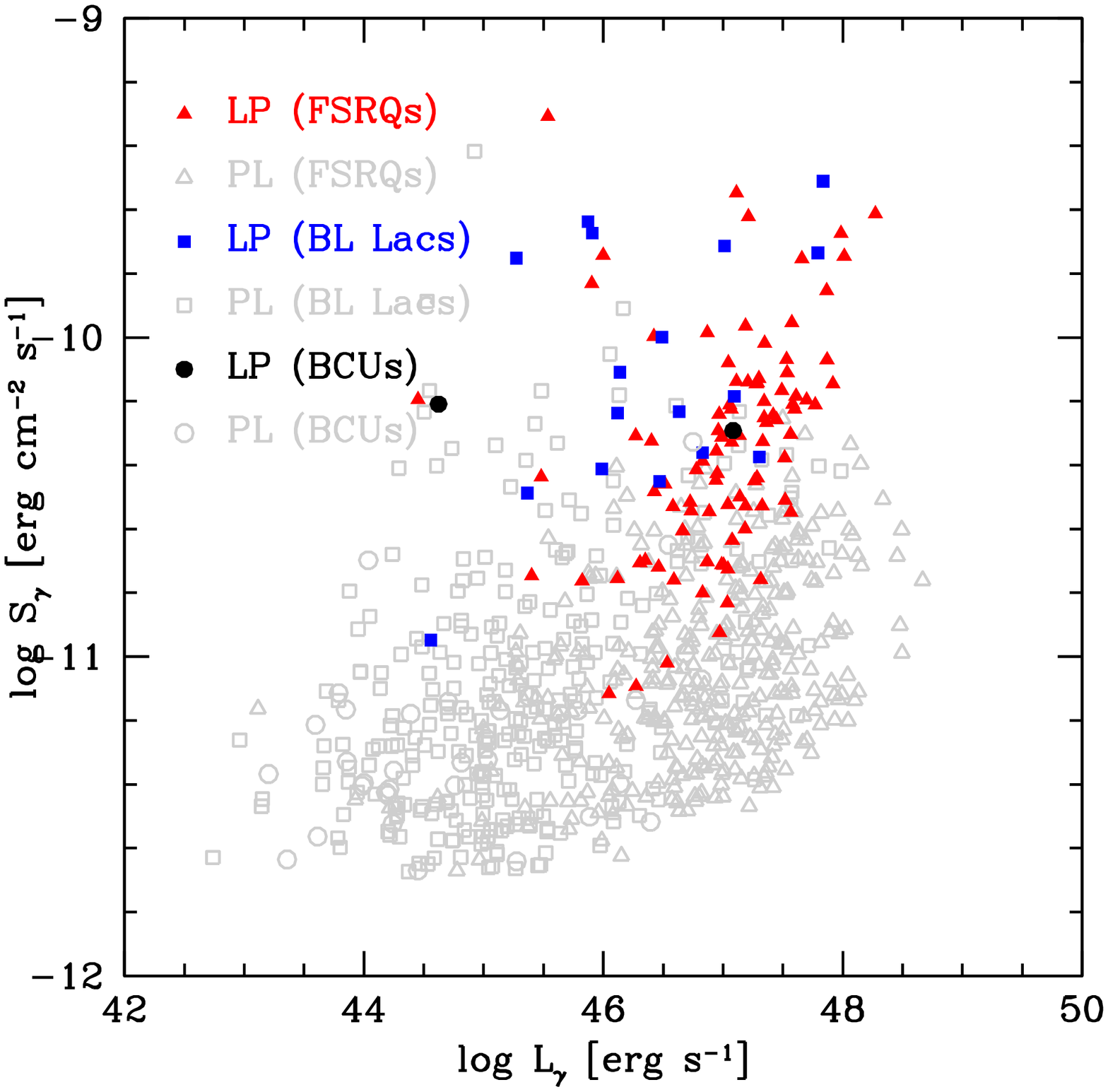}
   \includegraphics[height=6cm,width=7.2cm]  {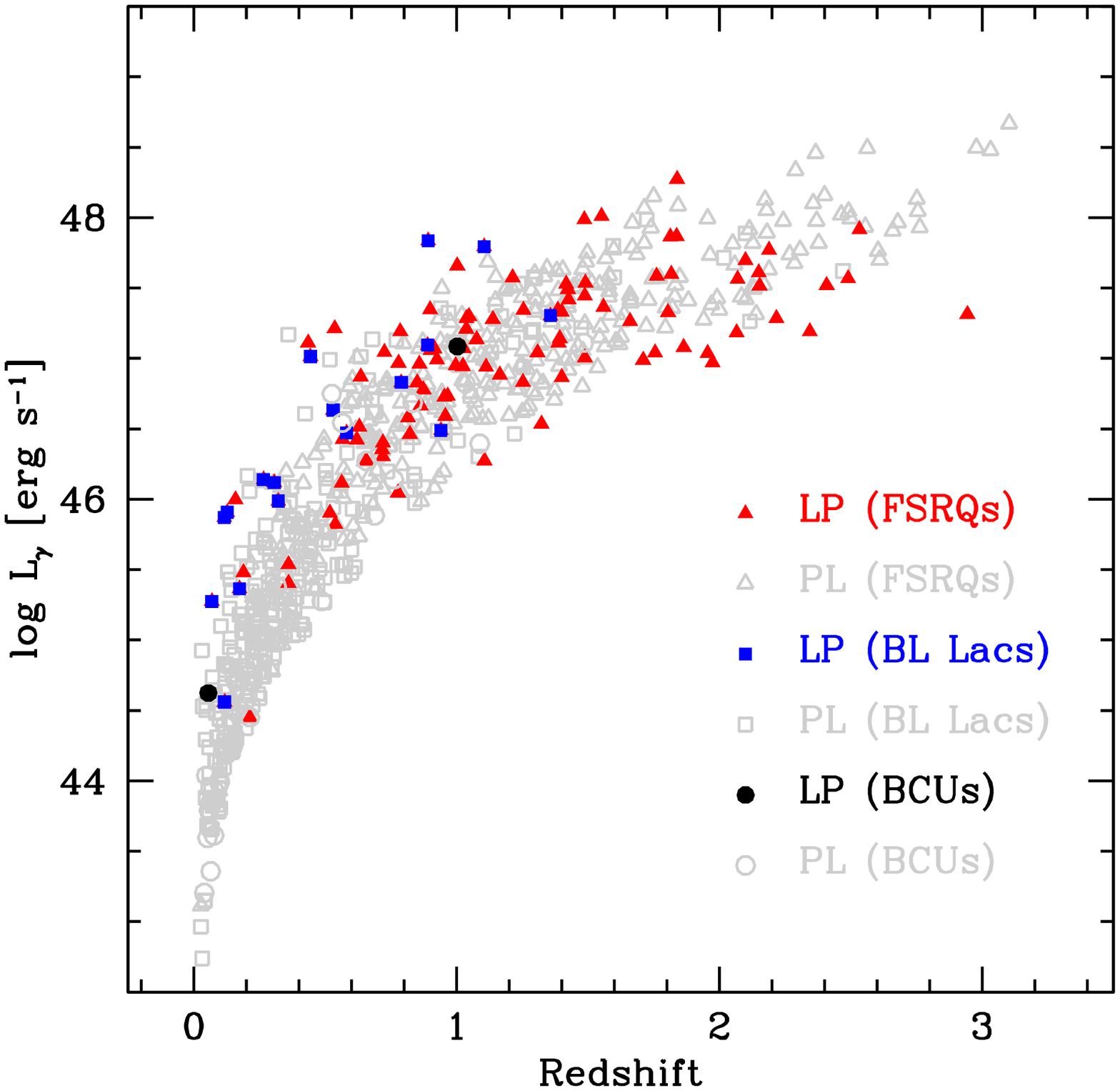}
   \caption{The correlations between variability index, energy flux (${\rm S}_{\gamma}$), gamma-ray luminosity (${\rm L}_{\gamma}$) and redshift for blazars.
   The red solid triangles, blue squares and black points indicate the observational data for FSRQs, BL Lacs and BCU blazars with the LP shape $\gamma$-ray spectrum; 
   The grey empty triangles, squares and points represent the observational data for FSRQs, BL Lacs and BCU blazars with the PL shape $\gamma$-ray spectrum. 
   }\label{sub:fig1}
  \end{figure*}

\begin{figure*}
   \centering
   \includegraphics[height=6cm,width=7.2cm]  {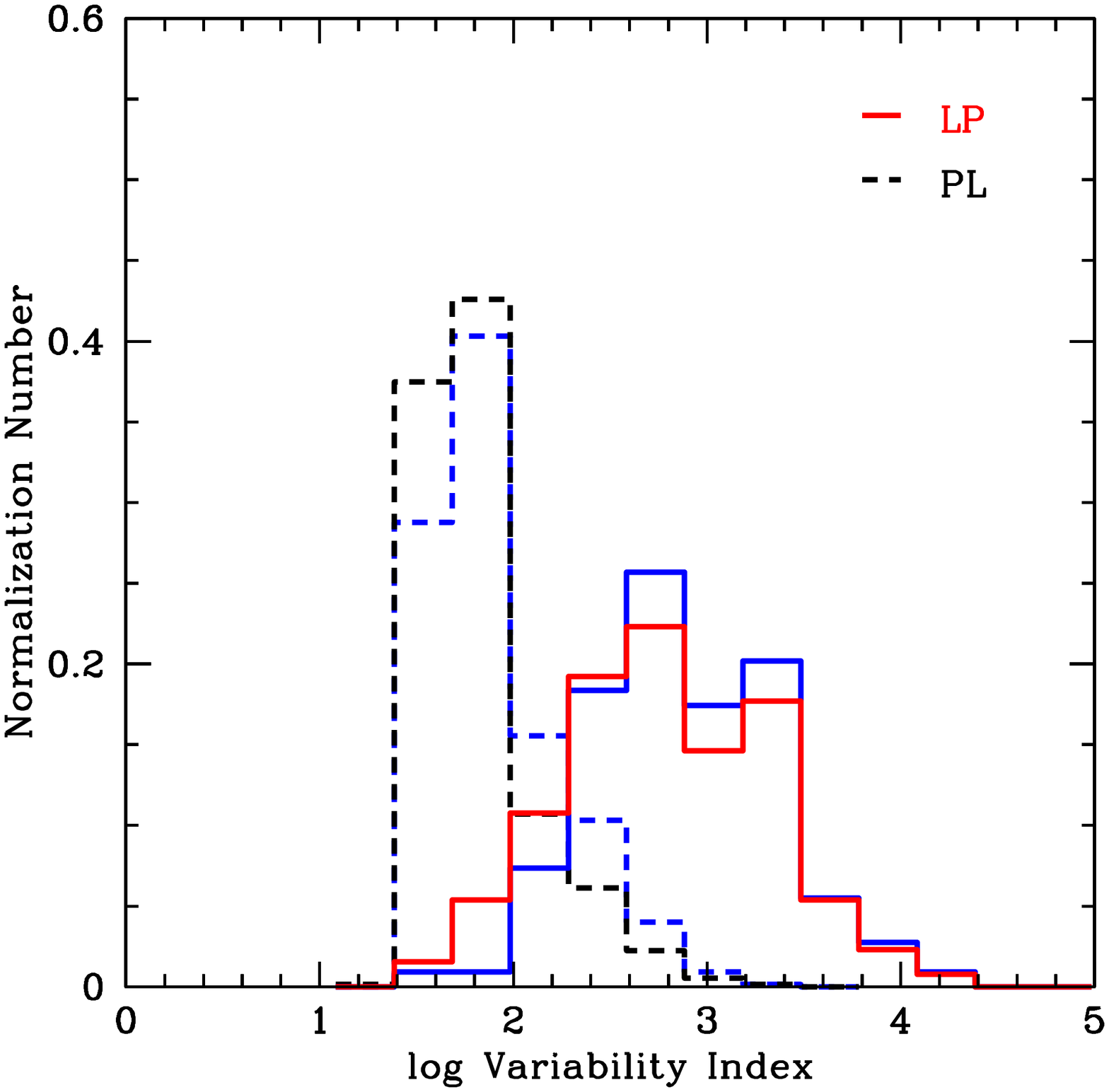}
   \includegraphics[height=6cm,width=7.2cm]  {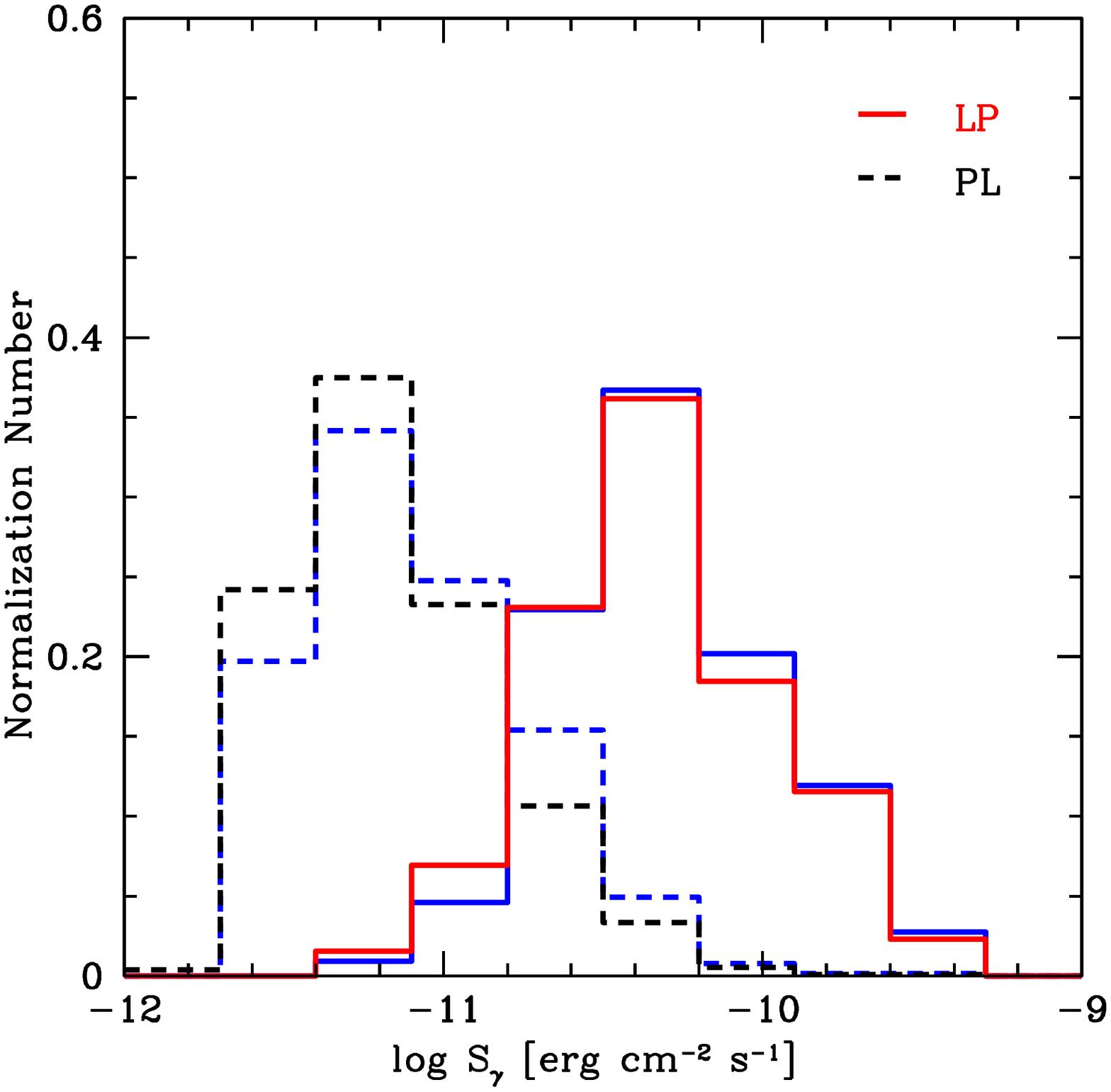}
   \includegraphics[height=6cm,width=7.2cm]  {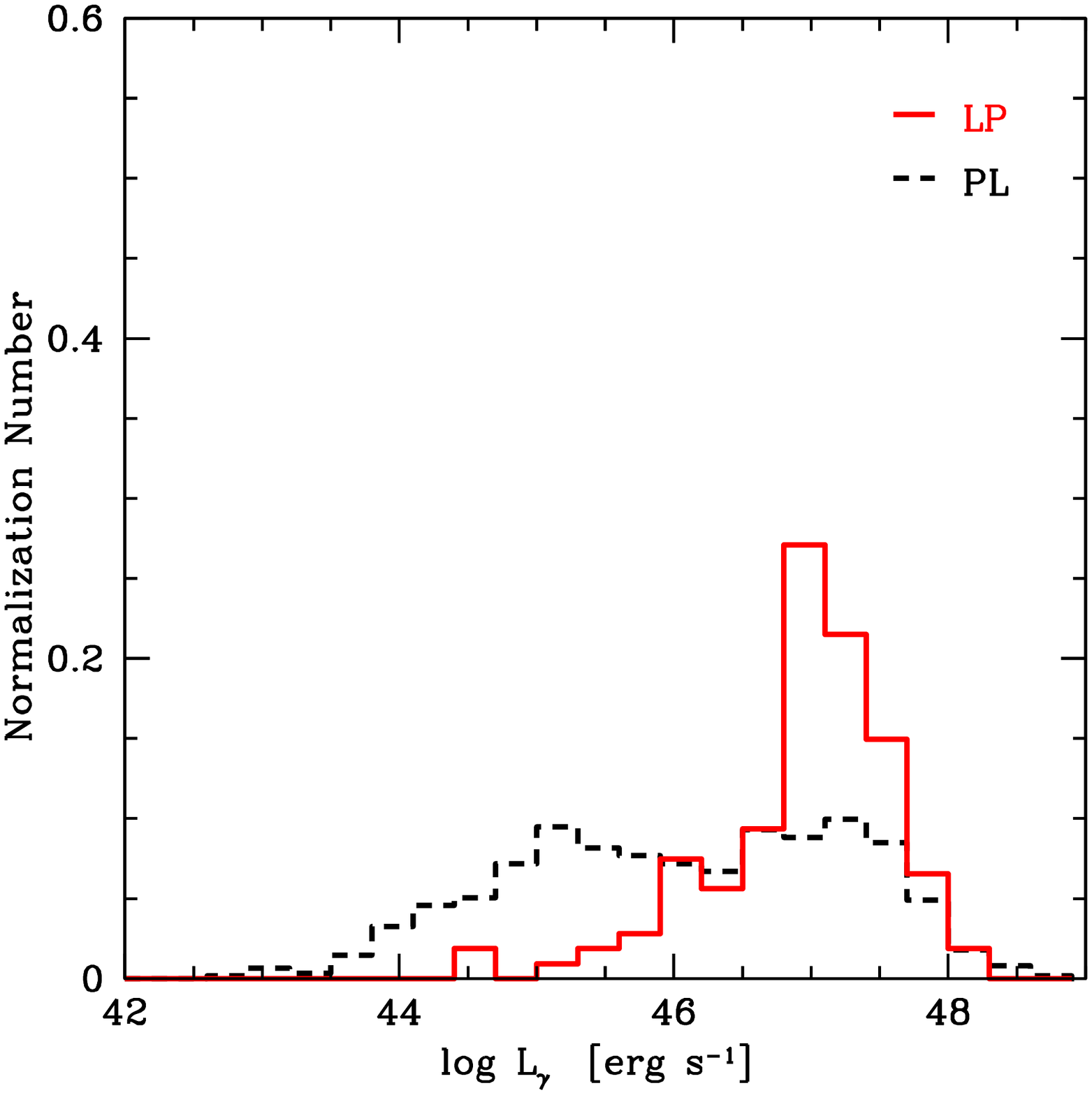}
   \includegraphics[height=6cm,width=7.2cm]  {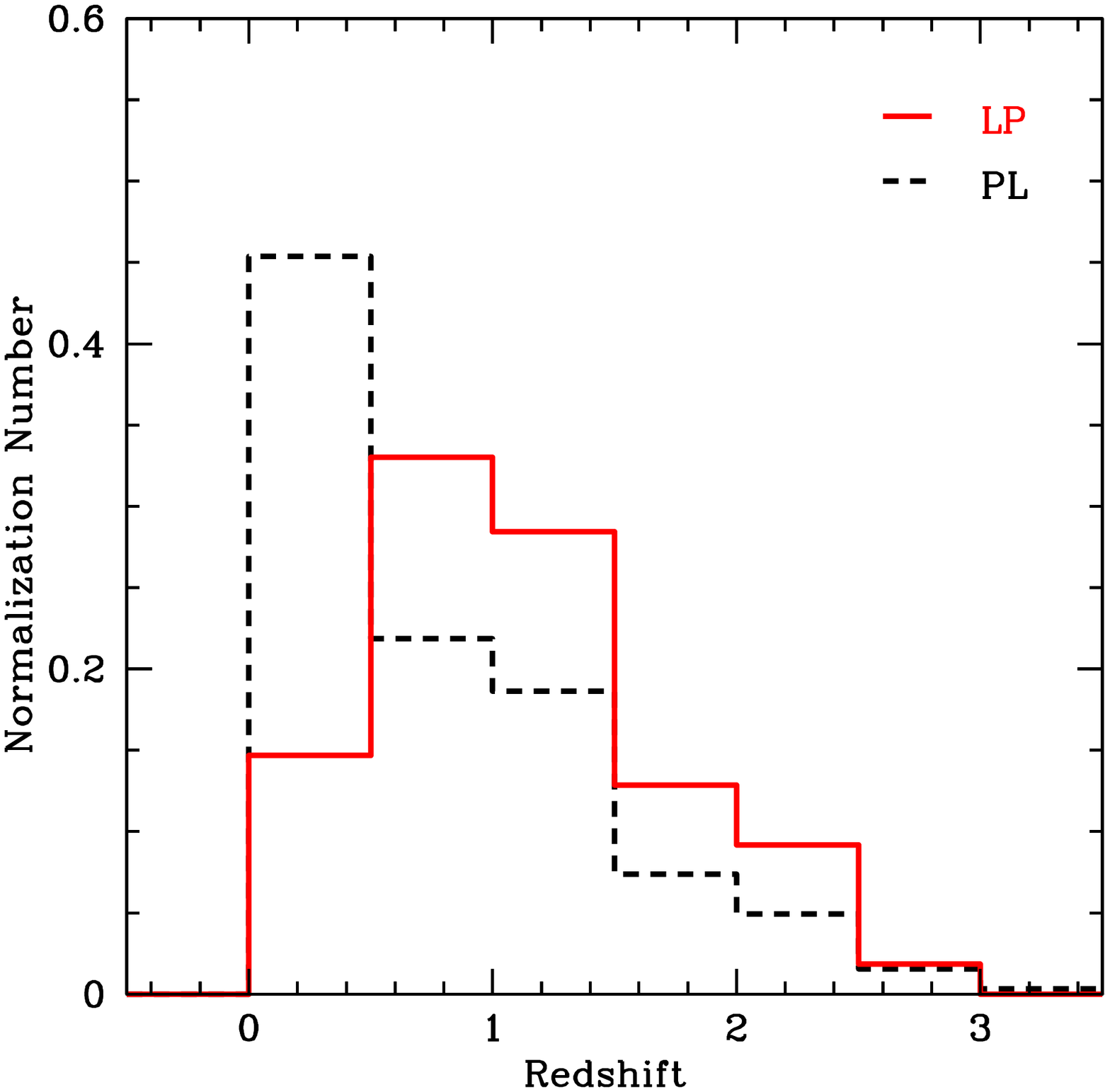}
   \caption{The normalized histograms for the variability index, $\gamma$-ray flux, $\gamma$-ray luminosity and redshift for the selected sources, where solid and dashed lines represent of LP and PL shape of $\gamma$-ray spectrum respectively. {In the up panels, the blue lines and dash lines indicate the results for the sources with known redshift.}
   }\label{sub:fig2}
  \end{figure*}

  We present the statistical results in Figures  \ref{sub:fig1}  and  \ref{sub:fig2} , where the correlations and histograms are shown for both PL and LP blazars. From the correlations of variability index-${\rm S}_{\gamma}$ and variability index-${\rm L}_{\gamma}$, we find that the LP sources normally have higher $\gamma$-ray flux and variability index compared the PL sources (top left panel of Figure \ref{sub:fig1}), where the variability index and ${\rm S}_{\gamma}$ show a positive correlation with a Pearson correlation coefficients of 0.81. The histograms of both variability index and $\gamma$-ray flux are shown evident bimodal distributions for LP and PL sources (top panels in Figure \ref{sub:fig2}), where the KMM-test (see \citealt{1994AJ....108.2348A} for the details and references therein) strongly reject that LP and PL sources follow a unimodal distribution (probabilities are $1.0\times10^{-\infty}$ and $1.25\times10^{-33}$  for variability index and ${\rm S}_{\gamma}$ respectively). {The bimodal distributions remain unchanged, even only consider the sources with known redshift (the blue lines and dash lines in top panels in Figure \ref{sub:fig2}}).

From the correlations of Variability index-redshift and ${\rm S}_{\gamma}$-redshift, we find that both LP and PL sources do not well correlated with the redshift. The KMM-test also show that there is roughly no difference for LP and PL sources in the histogram of redshift, where the probability is $p=0.995$ and reject the bimodal distribution of redshift for these two populations.  In the correlations of $\gamma$-ray flux-redshift and $\gamma$-ray luminosity-redshift, we can find that there are {no differences}  in the distributions of redshift and  $\gamma$-ray luminosity for LP and PL sources even though LP sources normally have the higher fluxes compared the PL sources. We also show the histogram of $\gamma$-ray luminosity for both LP and PL sources in bottom-right panel of Figure \ref{sub:fig2}, where these two populations should follow a unimodal distribution with a KMM-test probability of $p=0.999$.   In the correlation of $\gamma$-ray flux-$\gamma$-ray luminosity, we find that the bright sources with higher $\gamma$-ray flux show LP shape $\gamma$-ray spectrum {(e.g. \citealt{2012ApJS..199...31N}).}

\begin{table}
	\centering
	\caption{Probabilities of KMM-test}
	\label{tab:kmm}
	\begin{tabular}{ccccccc}  
		\hline
{probabilities} &    {$\rm S_{\gamma}$}  &   {VI}   &{z}  &   {${\rm L}_{\gamma}$}  \\
\hline
 p     	&	$1.0\times10^{-\infty}$  & $1.25\times10^{-33}$	&	0.995	&	0.999	       \\
\hline
	\end{tabular}\\
\end{table}

\section{Discussions ang Conclusions} \label{sec:conclusion}

Based on the 3LAC Clean Sample, we compile to 1419 Fermi blazars with 4 different parameters: variability index,  $\gamma$-ray energy flux, $\gamma$-ray luminosity and redshift. We explore the possible differences between LP and PL blazars based on the simple statistical analyses on these parameters. We find that the distributions of variability index and $\gamma$-ray flux are evidently different for the subsamples of LP and PL blazars (Figures \ref{sub:fig1}  and  \ref{sub:fig2}). However, we don't find evident differences in the distributions of $\gamma$-ray luminosity and redshift for LP and PL blazars. Our above conclusions remain unchanged for different optical classes (e.g., for the subsample of BL Lacs and FSRQs),
when the 402 BCU blazars  are excluded.

The unimodal distribution of $\gamma$-ray luminosity and bimodal distributions of variability index and $\gamma$-ray flux for LP and PL blazars may suggest that these two type of difference may be are observational effect and the $\gamma$-ray spectrum of all sources may be curved if we can detect more $\gamma$-ray photons. The $\gamma$-ray spectral shape of Fermi blazars can shed light on the possible intrinsic physics on the jet physics. For example, the emitting electron energy distribution (EED) can be determined by the modeling the SED of blazars {(e.g., \citealt{1986ApJ...308...78L,2007A&A...466..521T,2009A&A...501..879T,2013ApJ...765..122Y,{2013ApJ...771L...4C},{2011AcASn..52..357K,2012ChA&A..36..115K},{2010ApJ...721.1425A},2009A&A...504..821P,2004A&A...413..489M,{2004A&A...422..103M},2006A&A...448..861M,2014ApJ...788..179C})}. The form of the EED can give further information on the acceleration and cooling processes. First-order Fermi acceleration (shock acceleration) can naturally reproduce the PL EED (e.g., \citealt{{1983RPPh...46..973D},{1986MNRAS.223..353D},{1977ICRC...11..132A},{1999A&A...347..370D},{1977SPhD...22..327K},{1978MNRAS.182..147B},{1978MNRAS.182..443B},{1978ApJ...221L..29B},{1998A&A...333..452K}}). Second-order Fermi particle acceleration (stochastic acceleration) can form LP EED in the case of the acceleration process dominating over the radiative cooling {(e.g., \citealt{2004A&A...413..489M,2004A&A...422..103M,2006A&A...448..861M,2006ApJ...647..539B}; \citealt{2011ApJ...739...66T})}. If there is no acceleration process in the emitting region, the cooled EED is the broken PL shape {(e.g., \citealt{1999MNRAS.306..551C})}. Furthermore, it is found that the curvature of the spectrum may related to the peak frequency of the first hump in the SED of blazars, which seem to support the stochastic particle accelerations {(e.g., \citealt{2014ApJ...788..179C})}.
The intrinsic curved $\gamma$-ray spectrum may be  induced  by the curved EED coming from the stochastic acceleration process, or stochastic dominating acceleration process (e.g., \citealt{2018ApJ...853....6L}).
It should be noted that the extragalactic background light (EBL) might play a role in the curvature of the gamma-ray spectra altering the intrinsic blazars spectral shape. This will impact the conclusions on the intrinsic particle distribution in blazar jets (e.g., \citealt{2015MNRAS.450.4399A}). 

It is also important to note that the BCU blazars and/or without distance information might be biased towards BL Lac objects as the absence of spectral line features is a major hurdle in their identification and distance measurements. This is mandatory to exclude selection effects that can enter the statistical analyses of the parameter spaces thereby affecting conclusions on the jet physics of the Fermi-LAT blazars. The fainter and brighter sets of sources follow broader and more overlapping ranges in luminosity and redshift than they do in their flux. Such analysis would require issues of redshift completeness to be addressed , and the application of sophisticated statistical tools.  

In this work, in addition, we merely takes a published Fermi catalogue and makes various X-Y and histogram plots, adding no external data,  and have not done any of the fittings ourselves; only based on these data obtained from Fermi catalogue, for the limited  parameters, via a simple statistical Analysis (e.g.,bimodality test), to explore the intrinsic gamma-ray spectrum of Fermi blazar by compared the LP and PL blazars, which suggest a obvious inference that the fainter sources could be more complex in their properties --- more complex analysis is inhibited by the fact they are fainter sources. This result could be some bias, due to the simple statistical approach,  the limited sample  and the source of the sample data (obtained from the Fermi catalogue fitted by the Fermi team). The more preferable statistical approach, the big and completeness sample data, the more parameters (e.g., the mass of central black hole, and/or the polarization information of the jet radiation, etc.),  and the sophisticated fitting analysis are needed to further address the intrinsic jet physics of Fermi-LAT blazars.

\normalem
\begin{acknowledgements}

We thank the anonymous referee for very constructive and helpful comments and suggestions, which greatly helped us to improve our paper. 
This work is supported by the National Natural Science Foundation of China  (NSFC: 11763005, 11622324, 11573009, 11763002, U1431111, and U1431126). 
This work is aslo supported by the Research Foundation for Advanced Talents of Liupanshui Normal University (LPSSYKYJJ201506), 
{the Open Fund of Guizhou Provincial Key Laboratory of Radio Astronomy and Data Processing,}
the Physical Electronic Key Discipline of Guizhou Province (ZDXK201535), 
the Natural Science Foundation of the Department of Education of Guizhou Province (QJHKYZ[2015]455) 
and the Research Foundation of Liupanshui Normal University (LPSSYDXS1514, LPSSY201401).

\end{acknowledgements}
  

\end{document}